\documentclass[10pt]{iopart}
\lccode`\3`\3
\lccode`\/`\/
\usepackage[
    bottom=1.1cm
]{geometry}
\usepackage[per-mode = symbol]{siunitx}
\usepackage{tikz}
\usepackage{circuitikz}
\usepackage{pgfplots}
\pgfplotsset{compat=1.18}
\usepackage{pgfplotstable}
\usepgfplotslibrary{colormaps} 
\usetikzlibrary{external}
\usetikzlibrary{pgfplots.colormaps} 
\usetikzlibrary{decorations.markings}
\usetikzlibrary{patterns}
\makeatletter
\tikzset{
  nomorepostactions/.code={\let\tikz@postactions=\pgfutil@empty},
  mymark/.style 2 args={decoration={markings,
    mark= between positions 0 and 1 step (1/25)*\pgfdecoratedpathlength with{%
        \tikzset{#2,every mark}\tikz@options
        \pgfuseplotmark{#1}%
      },  
    },
    postaction={decorate},
    /pgfplots/legend image post style={
        mark=#1,mark options={#2},every path/.append style={nomorepostactions}
    },
  },
}
\makeatother
\makeatletter
    \tikzset{scale line widths/.style={%
    /utils/exec=\pgfgettransformentries{\tmpa}{\tmpb}{\tmpc}{\tmpd}{\tmp}{\tmp}%
    \pgfmathsetmacro{\myJacobian}{sqrt(abs(\tmpa*\tmpd-\tmpb*\tmpc))}%
    \pgfmathsetlength\pgflinewidth{\myJacobian*0.4pt}%
    \def\tikz@semiaddlinewidth##1{\pgfmathsetmacro{\my@lw}{\myJacobian*##1}%
    \tikz@addoption{\pgfsetlinewidth{\my@lw pt}}\pgfmathsetlength\pgflinewidth{\my@lw pt}},%
    thin}}
\makeatother
\usepackage{graphicx}
\usepackage{xcolor}
\usepackage[backend=biber, style=ieee, eprint=false]{biblatex}
\AtEveryBibitem{%
  \ifentrytype{online}{%
  }{%
    \clearfield{howpublished}%
    \clearfield{url}%
    \clearfield{urldate}%
    \clearfield{issn}%
  }%
}
\usepackage[
    pdftitle={An Open-Source 3D FE Quench Simulation Tool for NI HTS Pancake Coils},
    pdfauthor={Sina Atalay},
    pdfkeywords={quench simulation, finite element method, open-source},
    colorlinks,
    citecolor=blue,
    urlcolor=black,
    linkcolor=blue
]{hyperref}
\usepackage{droidsansmono}
\usepackage{subcaption}
\DeclareCaptionFormat{iop-caption}{\raggedright \footnotesize \textbf{#1#2}#3}
\DeclareCaptionFormat{iop-subcaption}{\centering\footnotesize #1#2#3}
\definecolor{BackgroundColor}{RGB}{250, 250, 250}
\definecolor{YamlKeyColor}{RGB}{75, 105, 198}
\definecolor{YamlValueColor}{RGB}{68,140,39}
\definecolor{YamlCommentColor}{RGB}{150, 150, 150}
\definecolor{YamlColonColor}{RGB}{119, 119, 119}
\usepackage{listings}
\usepackage[framemethod=TikZ]{mdframed}
\newmdenv[
    outerlinewidth=0.3,
    outerlinecolor=YamlCommentColor,
    middlelinewidth=0,
    backgroundcolor=BackgroundColor,
    roundcorner=3pt,
    innerbottommargin=0pt,
    innertopmargin=0pt,
    innerleftmargin=7pt,
    innerrightmargin=7pt,
    skipbelow=-0pt
]{yaml_box}
\newcommand\YAMLcolonstyle{\color{YamlColonColor}\droidsansmono\scriptsize\mdseries}
\newcommand\YAMLkeystyle{\color{YamlKeyColor}\droidsansmono\scriptsize\bfseries}
\newcommand\YAMLvaluestyle{\color{YamlValueColor}\droidsansmono\scriptsize\mdseries}
\makeatletter
\newcommand\language@yaml{yaml}
\expandafter\expandafter\expandafter\lstdefinelanguage
\expandafter{\language@yaml}
{
  keywords={true,false,null,y,n},
  keywordstyle=\YAMLkeystyle,
  basicstyle=\YAMLkeystyle,
  sensitive=false,
  comment=[l]{\#},
  morecomment=[s]{/*}{*/},
  commentstyle=\color{YamlCommentColor},
  stringstyle=\YAMLvaluestyle,
  moredelim=[l][\color{orange}]{\&},
  moredelim=[l][\color{magenta}]{*},
  moredelim=**[il][\YAMLcolonstyle{:}\YAMLvaluestyle]{:},
  morestring=[b]',
  morestring=[b]",
  literate =    {---}{{\ProcessThreeDashes}}3
                {>}{{\textcolor{red}\textgreater}}1     
                {|}{{\textcolor{red}\textbar}}1 
}
\lst@AddToHook{EveryLine}{\ifx\lst@language\language@yaml\YAMLkeystyle\fi}
\makeatother
\newcommand\ProcessThreeDashes{\llap{\color{cyan}\mdseries-{-}-}}
\begin{document}

\title[An Open-Source 3D FE Quench Simulation Tool for NI HTS Pancake Coils]{An Open-Source 3D FE Quench Simulation Tool for No-Insulation HTS Pancake Coils}

\author{Sina Atalay\textsuperscript{1, 2}, Erik Schnaubelt\textsuperscript{1, 3}, Mariusz Wozniak\textsuperscript{1}, Julien Dular\textsuperscript{1}, Georgia Zachou\textsuperscript{1, 4}, Sebastian Schöps\textsuperscript{3}, and Arjan Verweij\textsuperscript{1}}

\address{\textsuperscript{1} CERN, Switzerland}
\address{\textsuperscript{2} Boğaziçi University, Türkiye}
\address{\textsuperscript{3} Technical University of Darmstadt, Germany}
\address{\textsuperscript{4} National Technical University of Athens, Greece}
\ead{sina.atalay@cern.ch}
\vspace{10pt}
\begin{indented}
\item[]\today
\end{indented}

\begin{abstract}
A new module, \texttt{Pancake3D}, of the open-source Finite Element Quench Simulator (\texttt{FiQuS}) has been developed in order to perform transient simulations of no-insulation (NI) high-temperature superconducting (HTS) pancake coils in 3-dimensions. \texttt{FiQuS}/\texttt{Pancake3D} can perform magnetodynamic or coupled magneto-thermal simulations. Thanks to the use of thin shell approximations, an $\vec{H}-\phi$ formulation, and anisotropic homogenization techniques, each turn can be resolved on the mesh level in an efficient and robust manner. \texttt{FiQuS}/\texttt{Pancake3D} relies on pre-formulated finite-element (FE) formulations and numerical approaches that are programmatically adjusted based on a text input file. In this paper, the functionalities and capabilities of \texttt{FiQuS}/\texttt{Pancake3D} are presented. The challenges of FE simulation of NI coils and how \texttt{FiQuS}/\texttt{Pancake3D} addresses them are explained. To highlight the functionalities, the results of a magneto-thermal analysis of a double pancake coil with 40 turns per pancake with local degradation of the critical current density are conducted and discussed. Furthermore, a parameter sweep of the size of this local degradation is presented. All the \texttt{FiQuS} input files for reproducing the simulations are provided.
\end{abstract}

%
\vspace{2pc}

\submitto{\SUST}
%
\maketitle
%
\ioptwocol
%
\newlength{\pgfPlotWidth}
\setlength{\pgfPlotWidth}{0.89\linewidth}
\newlength{\pgfPlotHeight}
\setlength{\pgfPlotHeight}{4.6cm}
\newlength{\pgfPlotSweepHeight}
\setlength{\pgfPlotSweepHeight}{7.7cm}
\newcommand{\pgfPlotXTicks}{{0, 20, 40, 60, 80, 100}}
\newcommand{\pgfPlotXLabel}{Time (s)}

\pgfplotscreateplotcyclelist{mycyclelist}{%
  {index of colormap=17 of viridis, solid},%
  {index of colormap=16 of viridis, solid},%
  {index of colormap=15 of viridis, solid},%
  {index of colormap=13 of viridis, solid},%
  {index of colormap=10 of viridis, solid},%
  {index of colormap=8 of viridis, solid},%
  {index of colormap=6 of viridis, solid},%
  {index of colormap=3 of viridis, solid},%
  {index of colormap=0 of viridis, solid},%
  {black, solid}
}

\pgfplotsset{
    cycle list name=mycyclelist,
    enlarge x limits=false,
    every axis/.append style={
        line width=1pt
    },
    legend style={font=\footnotesize},
    label style={font=\footnotesize},
    tick label style={font=\footnotesize},
    grid=major,
    xlabel style = {
        yshift = 3pt
    },
    ylabel style = {
        yshift = -3pt
    }
}

\section{Introduction}\label{sec:introduction}
Quench detection and protection of high-field, high-temperature superconducting (HTS) magnets are crucial parts of the development of HTS technology. The quench behaviour of HTS magnets differs from low-temperature superconducting (LTS) magnets. HTS magnets have lower normal zone propagation velocity \cite{Tsukamoto2014}, and the critical current density of an HTS tape strongly depends on the direction of the magnetic field \cite{Ren2020}. Consequently, quench protection of HTS magnets is regarded as one of the challenging technical issues for HTS technology.

One of the propositions to protect HTS pancake coils is to allow current to divert around the quenched segments of the coil by winding HTS coated conductor (CC) without electrical insulation between the turns, i.e., no-insulation (NI) pancake coils \cite{Hahn2011}. NI pancake coils have attracted considerable research interest, and the design was demonstrated to be stable and effective in a range of high magnetic field applications \cite{Hahn2019, Yoon2016, Weijers2014, Liu_2020}. However, a few challenges posed by NI pancake coils have also been identified, such as overstrains caused by radial currents during quench \cite{Hahn2018}. Therefore, numerical simulations of the quench behaviour of NI pancake coils are essential.

However, the finite-element (FE) simulation of NI pancake coils is challenging for several reasons:
\begin{itemize}
    \item A quench is intrinsically a local effect, and the detailed study of the magneto-thermal transients requires a three-dimensional (3D) model, where each turn is resolved.
    \item The contact layer between the turns of a pancake coil is much thinner than the total thickness of the HTS CC. Therefore, meshing the contact layer in a classical way leads to either a low-quality mesh or a very fine mesh.
    \item HTS CCs consist of more than one material with highly non-linear properties, and homogenization methods are commonly preferred as the volumetric resolution of all layers is computationally infeasible for 3D simulations. The electrical resistivity of the HTS layer is a function of the current density in the layer. For this reason, homogenization requires an iterative approach to resolve the current sharing between HTS and other layers \cite{Bortot2020}. Furthermore, due to the layered structure of the HTS CC, the homogenization of the electrical resistivity and thermal conductivity should be considered as anisotropic. Finally, accessing material data and integrating it efficiently within the simulation software alongside homogenization techniques can be cumbersome.
    \item Implementing a comprehensive, robust, and efficient FE model of NI pancake coils requires a good understanding of FE formulations and an FE software package.
\end{itemize}

In this context, we present a new module of the open-source \cite{fiqus} and parametric Finite Element Quench Simulator (\texttt{FiQuS} \cite{Vitrano2023}), which is developed as part of STEAM (Simulation of Transient Effects in Accelerator Magnets \cite{steam}) framework at CERN. The new \texttt{Pancake3D} module can generate geometry, generate mesh, solve and postprocess magnetodynamic or coupled magneto-thermal FE simulation of a single or stack of multiple NI pancake coils. 

\texttt{FiQuS} is a Python application that utilizes Gmsh \cite{gmsh} for geometry and mesh generation and uses GetDP \cite{getdp} as the finite-element solver. Our main objective is to provide access to simulations of superconducting magnets to all researchers, regardless of their level of expertise in modelling and simulation. The software design of \texttt{FiQuS} provides a user interface that separates the coil design and the powering details from the numerical computing aspect. This separation allows users to describe and simulate an NI pancake coil setup by simply interacting with a text-based input file without worrying about mathematical formulations.

This paper presents an overview of \texttt{FiQuS}/\texttt{Pan\-cake3D}'s principles and functionalities, along with examples. The features and capabilities of the user interface, geometry generator, mesh generator, solver, and postprocessor are reviewed in \sref{sec:features}. The modelling approaches employed, like thin-shell approximations, STEAM material library integration, and homogenization methods, are also presented in \sref{sec:features}. Then, in \sref{sec:results}, results for a magneto-thermal analysis of a double pancake coil with a critical current density degradation in a segment of the HTS CC are presented. Also, a parameter sweep of the size of the degradation is performed and presented. Finally, conclusions are drawn in \sref{sec:conclusions}.

\section{Features and Capabilities}\label{sec:features}
This section describes the functionalities and capabilities of \texttt{FiQuS}/\texttt{Pan\-cake3D}, along with implementation details. \Sref{sec:yaml} presents the user interface and its usage. \Sref{sec:geometry} then reviews the techniques used to generate geometries efficiently and summarizes implemented features. \Sref{sec:mesh} demonstrates the capabilities of the mesh generator. \Sref{sec:solver} discusses the numerical methods used to conduct simulations and the available solver features. Finally, \sref{sec:postprocessor} showcases the postprocessing capabilities.

\subsection{YAML Input File}\label{sec:yaml}
The only user interface of \texttt{FiQuS} is a text-based YAML \cite{Ben2009} input file. For each simulation, there is one corresponding input file. The text-based input file makes it easy to version-control simulations and aids their reproducibility. The input files for the simulations presented in \sref{sec:results} are provided \cite{referenceInputs} so that the same results can be easily reproduced.

The YAML input file is a concise description of the model. It consists of four main parts: \textit{geometry}, \textit{mesh}, \textit{solve}, and \textit{postprocess}. With this approach, users can concentrate on the coil setup instead of the numerical methods that \texttt{FiQuS} handles in the background. For example, the input to generate a geometry of a double pancake coil with 20 turns per pancake is shown in \fref{fig:yaml_showcase}.

\begin{figure}[tbh]
    \centering
    \input{figures/features_and_capabilities/yaml_showcase}
    \caption{An example \textit{geometry} section of a YAML input file that generates a geometry of a double pancake coil with 20 turns per pancake as shown in \fref{fig:geometry_showcase}.}
    \label{fig:yaml_showcase}
\end{figure}

To simplify the usage of the YAML input files, \texttt{FiQuS} provides a standardized schema that specifies and documents the structure of the input format, using JSON Schema \cite{Pezoa2016}. Moreover, all the input data is validated initially to minimize the risk of encountering unexpected issues at the later stages. Warnings will be given if wrong or unfeasible inputs are provided, such as a negative inner radius of windings or unrealistic choices for the mesh. 

\subsection{Geometry Generator}\label{sec:geometry}
\begin{figure}[tbh]
    \centering
    \begin{tikzpicture}
    \begin{axis}[
        width=0.86\pgfPlotWidth,
        axis equal image,
        hide axis,
        enlargelimits=false,
        ticks=none,
        legend columns=-1,
        legend style={
            at={(0.5, -0.04)},
            anchor = north, 
            /tikz/every even column/.append style={column sep=0.3cm}
        }
    ] 
            \definecolor{terminalsColor}{RGB}{214,175,160}
            \definecolor{windingColor}{RGB}{154,148,145}
            \definecolor{gapColor}{RGB}{0,0,0}
            
            \addlegendimage{fill=terminalsColor, area legend}
            \addlegendimage{fill=windingColor, area legend}
            \addlegendimage{fill=gapColor, area legend}
            
            \addlegendentry{\footnotesize Terminals}
            \addlegendentry{\footnotesize Windings}
            \addlegendentry{\footnotesize Air Gap}

            \addplot graphics[xmin=0,xmax=1757,ymin=0,ymax=1738] {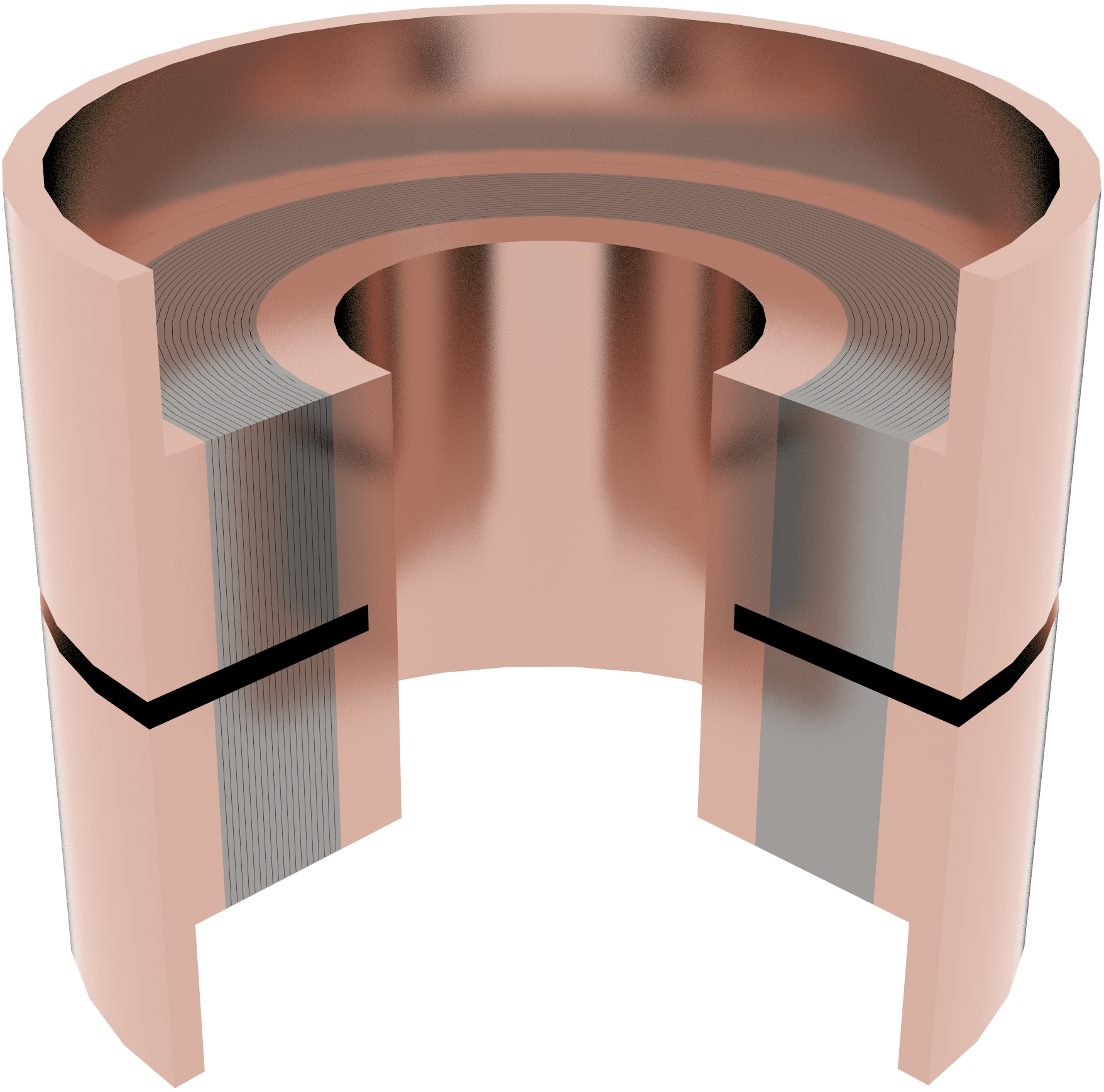};
            
    \end{axis}
\end{tikzpicture}%
    \caption{Clipped view of a double pancake coil geometry generated from the input shown in \fref{fig:yaml_showcase}.}
    \label{fig:geometry_showcase}
\end{figure}

\texttt{FiQuS}/\texttt{Pan\-cake3D} is capable of generating fully parametric pancake coil geometries, using Open CASCADE Technology (OCCT) \cite{occt} via Gmsh \cite{gmsh}. Pancake coils are generated as conformal spiral windings with two terminal blocks that are conformal to the inner and outer surfaces of the windings. An air region conformal to the coil geometry is also generated. Terminal blocks are extended axially to the end of the air region as tubes. A stack of multiple pancake coils can be generated with alternating connections through inner and outer terminal blocks. All the pancake coils in the stack must have the same inner radius, outer radius, and number of turns. Such geometries are generated very efficiently. For example, the geometry of a stack of 10 pancake coils, each with 60 turns, including the air volume that surrounds the coils, can be generated and saved in under 2 minutes using an Intel Core i7-3770 and CPython \cite{python} via Gmsh Python API. The significant performance gain is achieved by avoiding Boolean operations on volumes by creating boundary surfaces and constructing the volumes from closed surface loops. \Fref{fig:geometry_showcase} shows a double pancake coil geometry produced from the input in \fref{fig:yaml_showcase}.

\Fref{fig:yaml_showcase} summarizes all available inputs for the geometry generator. The contact layer between the turns can be modelled as either a 2D surface or a 3D volume. Also, an additional and optional air shell volume that surrounds the main air volume can be generated to perform shell transformation to model the infinite space (see \sref{sec:solver}).

\subsection{Mesh Generator}\label{sec:mesh}
\texttt{FiQuS}/\texttt{Pancake3D} uses the open-source mesh generator Gmsh to mesh the generated pancake coil geometry. The mesh of each region (windings, terminals, and air) can be controlled separately, thereby providing the flexibility needed to generate the appropriate mesh that is suitable for a given problem.

\begin{figure}[tbh]
    \centering
    \includegraphics[width=0.62\linewidth]{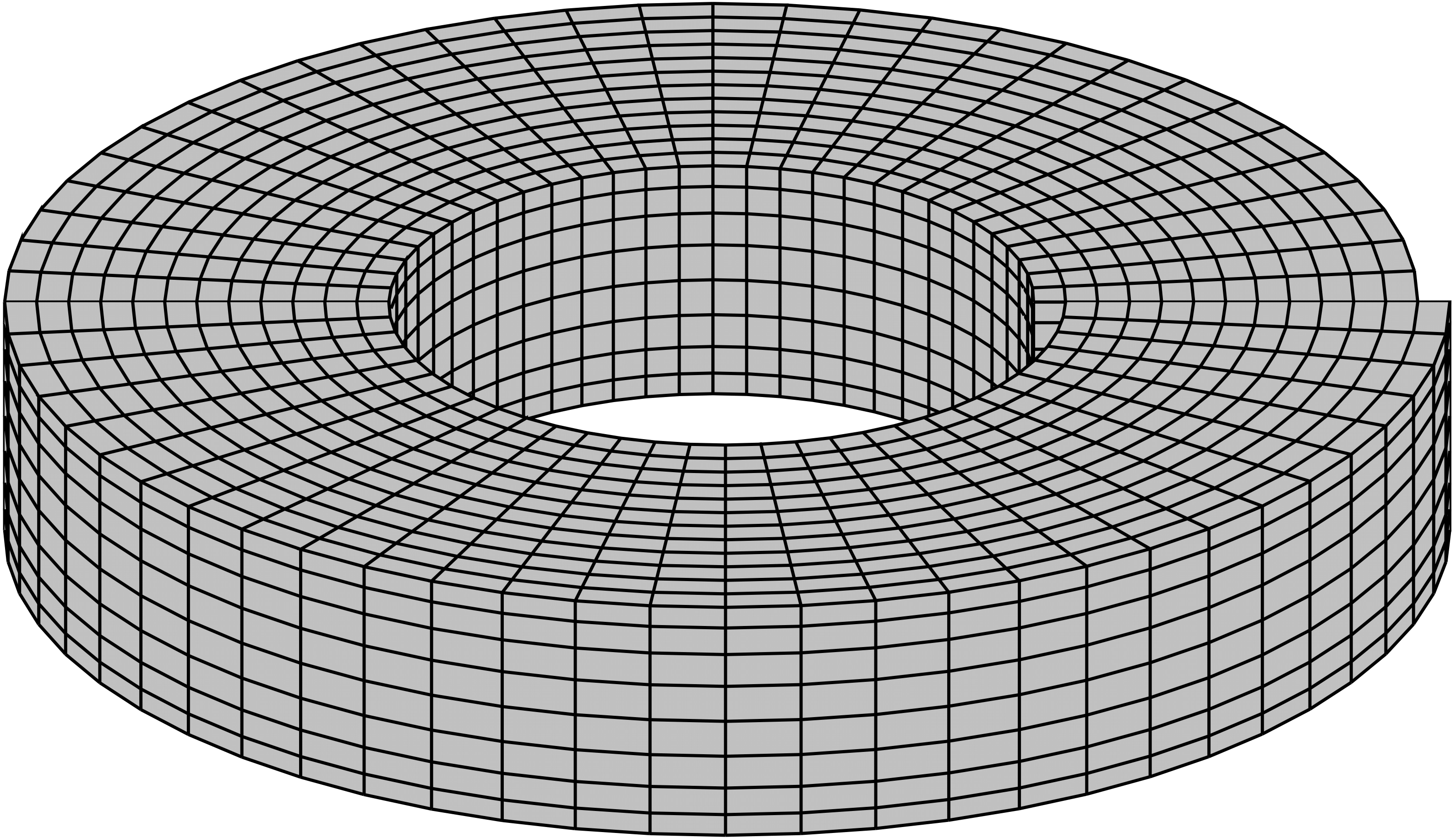}
    \caption{A pancake coil winding mesh with hexahedral elements and a 2D contact layer between the turns from the input shown in \fref{fig:yaml_mesh_winding}.}
    \label{fig:mesh_winding}
\end{figure}

\Fref{fig:mesh_winding} shows an example of a winding mesh generated from the input shown in \fref{fig:yaml_mesh_winding}. For the windings of the pancake coil, users can select the number of azimuthal elements, axial elements, radial elements per turn, and one of three element types: tetrahedron, hexahedron, or prism. Moreover, the size of mesh elements in the axial direction does not have to be homogeneous and can be controlled with an "axial distribution coefficient" so that finer meshes at the edges can be achieved (see \fref{fig:mesh_winding} for an example). Finally, if the geometry consists of more than one pancake coil, inputs can be independently specified for each winding.

\begin{figure}[tbh]
    \centering
    \input{figures/features_and_capabilities/yaml_mesh_winding}
    \caption{An example \textit{mesh} section of a YAML input file that generates a winding mesh as shown in \fref{fig:mesh_winding}.}
    \label{fig:yaml_mesh_winding}
\end{figure}

\begin{figure}[tbh]
    \centering
    \begin{subfigure}{.47\linewidth}
      \centering
      \includegraphics[width=1\linewidth]{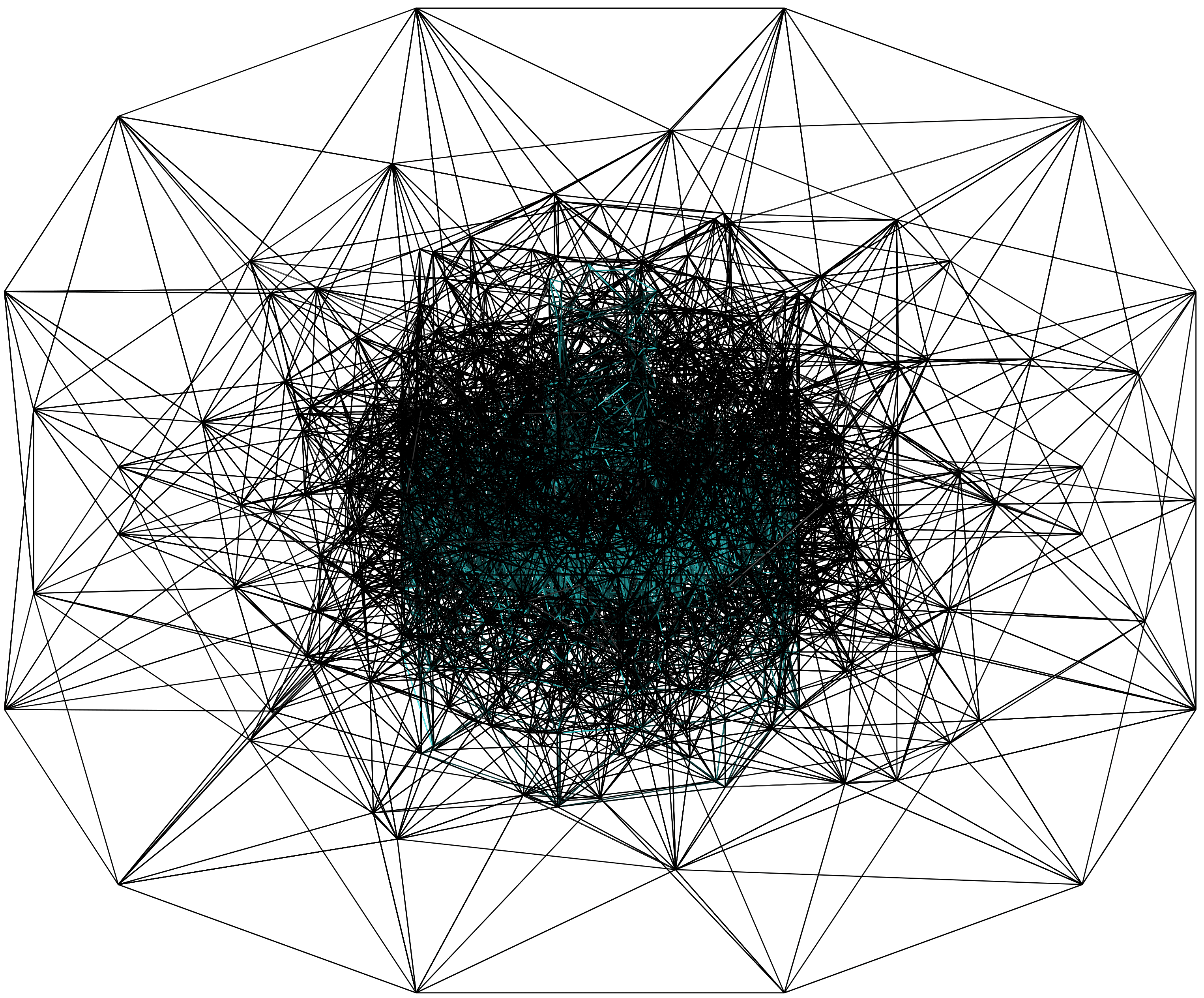}
      \caption{With unstructured air and terminal mesh.}
      \label{fig:sub1}
    \end{subfigure}%
    \hspace{0.03\linewidth}
    \begin{subfigure}{.47\linewidth}
      \centering
      \includegraphics[width=1\linewidth]{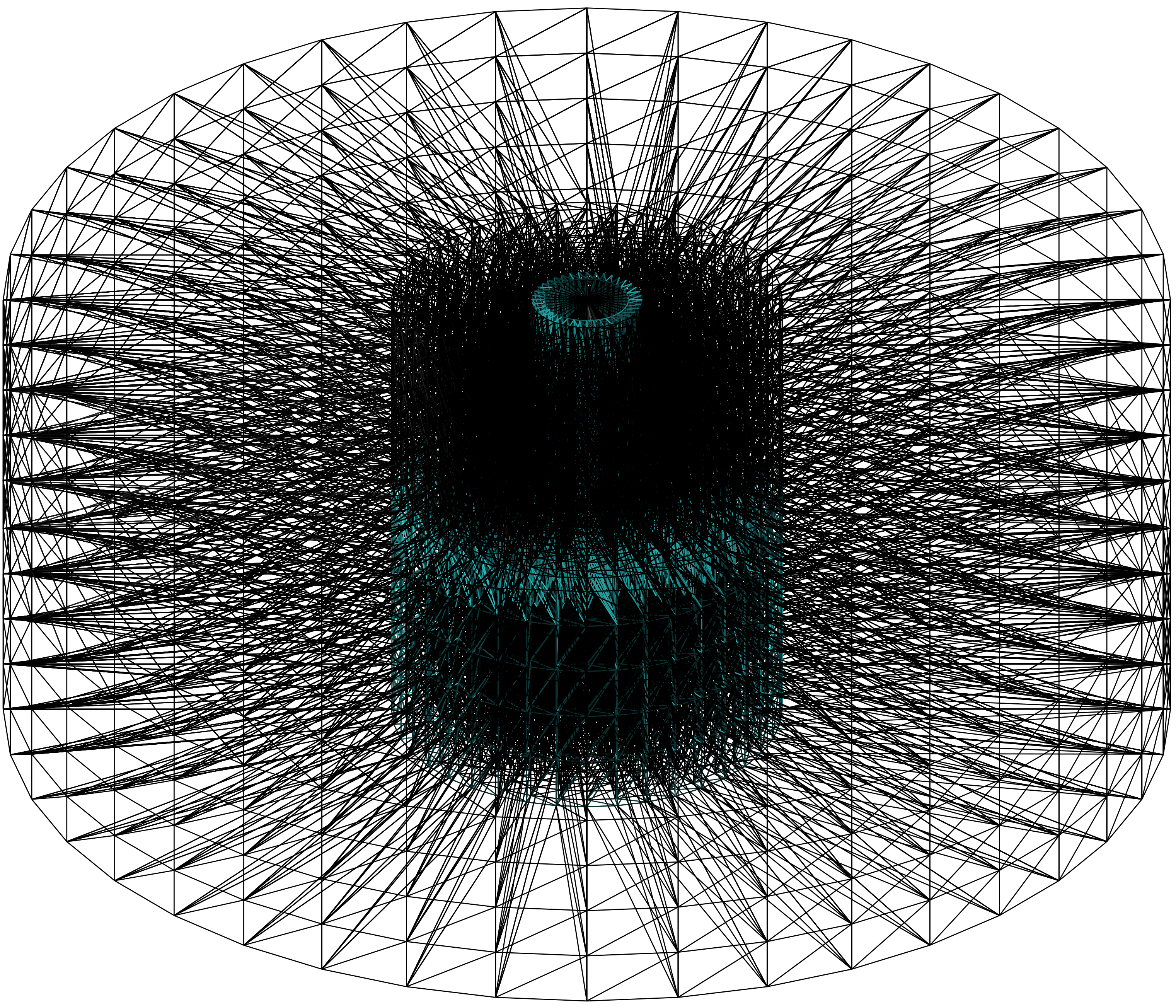}
      \caption{With structured air and terminal mesh.}
      \label{fig:sub2}
    \end{subfigure}
    \caption{Two examples of a complete pancake coils mesh, including air and terminals.}
    \label{fig:mesh_total}
\end{figure}

For terminal and air regions, both structured and unstructured meshes are supported. When using an unstructured mesh, users can specify maximum and minimum element size to generate finer meshes around the winding and coarser meshes farther away. This capability allows for a more precise analysis near the winding region without the need to uniformly refine the mesh. However, a structured mesh can also be achieved, and the radial element sizes of terminals and air can be specified. Two examples of a complete mesh, one with unstructured air mesh and the other with structured air mesh, are shown in \fref{fig:mesh_total}.

\subsection{Solver}\label{sec:solver}
\texttt{FiQuS} employs GetDP \cite{getdp} as the finite element solver and uses the STEAM material library \cite{steamMaterialLibrary} for material properties. GetDP is an open-source software environment that allows users to define general FE problems with mathematical expressions. \texttt{FiQuS}/\texttt{Pancake3D} is purpose-built on top of this general framework for quench simulations of NI pancake coils with the following solver features, which will be explained in detail afterwards:

\begin{itemize}
    \item Magnetodynamic or coupled magneto-thermal simulations can be conducted.
    \item Thin-shell approximations are supported for contact layers.
    \item A shell transformation can be used to model infinite space with finite elements.
    \item Temperature- and magnetic field-dependent material properties can be assigned to each region explicitly by specifying the material names or alternatively specifying constant values for electrical resistivity, thermal conductivity, and specific heat capacity.
    \item The stack of materials of an HTS CC can be specified with a list of material names with their relative thicknesses. Any number of materials can be stacked onto each other.
    \item Three different cooling conditions can be used.
    \item Local degradations of critical current (local defects) along the HTS CC can be introduced either by specifying the critical current as a function of the tape length or assigning a custom critical current density value to a specific range of turns.
\end{itemize}

\texttt{FiQuS}/\texttt{Pancake3D} is capable of conducting both magnetodynamic and coupled magneto-thermal simulations. For magnetodynamics, it uses an $\vec{H}-\phi$ magnetodynamic formulation with details in \cite{Schnaubelt2023Electromagnetic}. This formulation is an efficient and robust choice for systems of HTS without ferromagnetic materials \cite{Dular2020}. For coupled magneto-thermal simulations, alongside the $\vec{H}-\phi$ formulation, it solves the heat equation. The $\vec{H}-\phi$ formulation and the heat equation are coupled by resistive heating acting as a heat source as well as through the temperature- and field-dependent material properties. The mathematical formulation is found in \cite{Schnaubelt2023Coupled}.

One of the problems with the FE modelling of NI pancake coils is meshing the thin contact layer. The contact layer's thickness is commonly much smaller compared to the HTS CC's thickness. Therefore, meshing it with 3D elements leads to either an exceedingly high number of degrees of freedom or a low-quality mesh. \texttt{FiQuS}/\texttt{Pan\-cake3D} overcomes this problem by modelling the contact layer with 2D shell elements and enforcing appropriate interface conditions on the two sides of the layer that approximate the 3D layer. This method is called thin-shell approximation, and it has been successfully verified against solutions with 3D contact layers in thermal simulations \cite{Schnaubelt2023Thermal}, magnetodynamic simulations \cite{Schnaubelt2023Electromagnetic}, and coupled magneto-thermal simulations \cite{Schnaubelt2023Coupled}.

An option to approximate an open infinite space using a shell transformation method, provided by GetDP, for the magnetic field \cite{Dular1999} is also available. This option allows a smaller air mesh to be used while maintaining sufficient accuracy.

The integration with the STEAM material library makes it possible to assign material properties to each region by specifying material names. Users can also choose to use constant values for material properties. For instance, \fref{fig:yaml_material_showcase_0} shows an example of a YAML input file's \textit{solve} section that assigns properties of copper to the terminals and constant properties to the contact layer.

\begin{figure}[tbh]
    \centering
    \input{figures/features_and_capabilities/yaml_material_showcase_0}
    \caption{An example \textit{solve} section of a YAML input file that assigns material properties of the pancake coil's terminals and the contact layer.}
    \label{fig:yaml_material_showcase_0}
\end{figure}

Another difficulty of FE simulations of NI pancake coils is that HTS CCs consist of more than one material with nonlinear properties. Their material properties need to be approximated with homogenization since it is not computationally feasible to resolve each layer in a 3D mesh. The homogenization requires an iterative approach to resolve the current sharing between the HTS and other layers due to the $E-J$ power-law modelling of the HTS \cite{Bortot2020, Schnaubelt2023Coupled}. The homogenization leads to anisotropic properties. Indeed, as explained in \cite{Schnaubelt2023Coupled}, the effective electrical resistivity and thermal conductivity are different along the thickness, width and length directions of the CC. The resulting tensorial structure is handled automatically by means of local coordinate transformations. In the input file, the materials of the CC are specified as a list of material names with their relative thicknesses, as shown in \fref{fig:yaml_material_showcase}. Additionally, a metallic shunt layer at the narrow side of the CC can be specified in the input, which will be taken into account during the homogenization process as a parallel resistivity in the thickness direction of the CC. Alternatively, constant values can be used for the winding tape's material properties.

\begin{figure}[tbh]
    \centering
    \input{figures/features_and_capabilities/yaml_material_showcase}
    \caption{An example \textit{solve} section of a YAML input file that assigns material properties of a pancake coil's HTS CC.}
    \label{fig:yaml_material_showcase}
\end{figure}

Because the materials exhibit nonlinear behaviour, iterative techniques are necessary. \texttt{FiQuS}/\texttt{Pan\-cake3D} uses Picard (fixed point) iterations for all material parameters, except for the electrical resistivity of the CC, for which the Newton-Raphson method is used for linearization with respect to the current density, as proposed in \cite{Dular2020} for the $\vec{H}-\phi$ formulation. Several convergence criteria for the nonlinear solver can be specified in the input file. The convergence criteria can be based on physical quantities or on the vector of degrees of freedom. 

For cooling conditions, three options are offered: adiabatic, a cryocooler boundary condition imposed on top and bottom terminal surfaces, and fixed terminal surface temperatures. The adiabatic option prevents any thermal exchange through terminal surfaces so that all generated heat is deposited in the solids of the model. The cryocooler boundary condition uses functions from the STEAM material library specifying the cooling power as a function of temperature based on the load map provided by manufacturers. Finally, the top and bottom terminal surface temperatures can be fixed.

\texttt{FiQuS}/\texttt{Pan\-cake3D} has a feature that allows quench initiation with local defects along the HTS CC. Different material properties can be assigned to different segments of the HTS CC. To introduce local defects, users can specify the critical current as a function of tape length, or they can assign a custom critical current density to a specific segment specified with turn numbers (for example, from turn 10.2 to 11.6).

\texttt{FiQuS}/\texttt{Pancake3D} simulates a transient phenomenon. For time discretization and integration, users can choose either adaptive or fixed time stepping. Adaptive time stepping provides an efficient discretization of time, where small time steps are used only when necessary. The time step size varies according to the criteria given in the input file. Similar to the nonlinear solver's convergence criteria, multiple criteria can be specified based on physical quantities or on the vector of degrees of freedom. Users can also choose a fixed time stepping for the uniform discretization of time. For adaptive time stepping, the backward differentiation formula (BDF) of order 1 to 6 can be used for integration. For fixed time stepping, BDF of order 1 (i.e., the implicit Euler method) is used.

The above solver features are abstracted away from the user by using Jinja \cite{jinja} templating engine to create the input files for GetDP. GetDP does not have an application programming interface (API), and it requires a text-based input file that defines the whole FE problem in a special format (i.e., the .pro file). \texttt{FiQuS}/\texttt{Pan\-cake3D} automatically generates GetDP input files with Jinja for every simulation.

\subsection{Postprocessor}\label{sec:postprocessor}
\texttt{FiQuS}/\texttt{Pancake3D} saves all the outputs in text-based files after each simulation and provides a postprocessor that can transform the numerical data into figures. It has two main postprocessing features: time-series plots and a magnetic field on a cutting plane. Additionally, Gmsh's graphical user interface can be used to visualize the vector and scalar fields, along with Gmsh's postprocessing capabilities.

The postprocessor can plot any physical quantity, such as temperature, current density, resistivity, magnetic field, etc., at any model coordinate with respect to time. The probing point's coordinates can be specified either with XYZ coordinates or a turn number (possibly non-integer) with a pancake coil index. For example, \fref{fig:yaml_time_series_showcase} shows an example input to plot the magnitude of the current density at the location corresponding to turn 10.3 of the first pancake coil with respect to time, and \fref{fig:time_series_showcase} shows the result.

\begin{figure}[tbh]
    \centering
    \input{figures/features_and_capabilities/yaml_time_series_showcase}
    \caption{An example \textit{postprocess} section of a YAML input file that generates a time-series plot of the magnitude of the current density at turn 10.3 of the first pancake coil.}
    \label{fig:yaml_time_series_showcase}
\end{figure}

\begin{figure}[tbh]
    \centering
    \includegraphics[width=\linewidth]{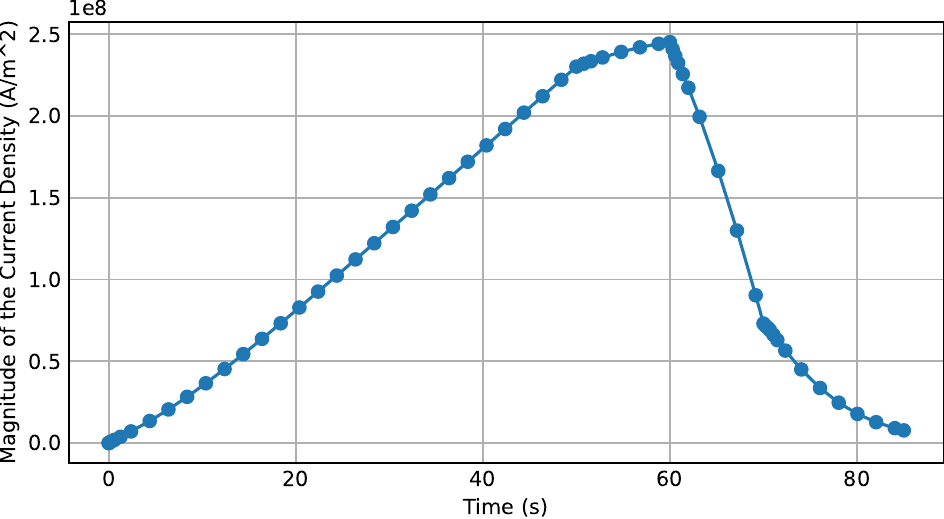}
    \caption{An example time-series plot of the magnitude of the current density at turn 10.3 of the first pancake coil generated by the postprocessor from the input shown in \fref{fig:yaml_time_series_showcase}.}
    \label{fig:time_series_showcase}
\end{figure}

\begin{figure}[tbh]
    \centering
    \includegraphics[width=\linewidth]{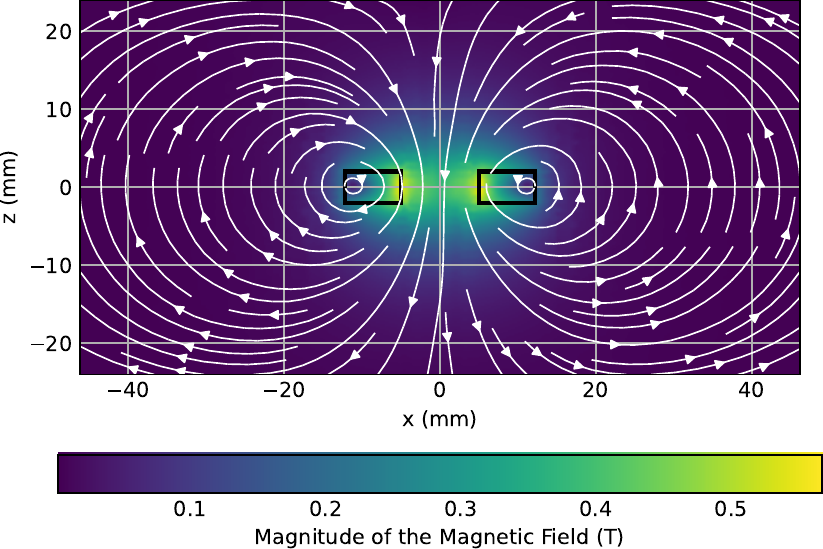}
    \caption{Magnetic field presentation on an XZ-plane generated by the postprocessor with the results obtained from a simulation with structured air mesh.}
    \label{fig:cutplane_showcase}
\end{figure}

The postprocessor can also present the magnetic field on a cutting plane as a 2D map with streamlines. Users can select any arbitrary cutting plane that passes through the origin by specifying the normal of the plane in the model's coordinate system, where the z-axis is oriented along the axial direction of the pancake coils. The magnitudes of the magnetic field are interpolated to obtain a smooth 2D map. \Fref{fig:cutplane_showcase} shows an example 2D magnetic field map obtained from a simulation with structured air mesh.

\begin{figure}[tbh]
    \centering
    \includegraphics[width=\linewidth]{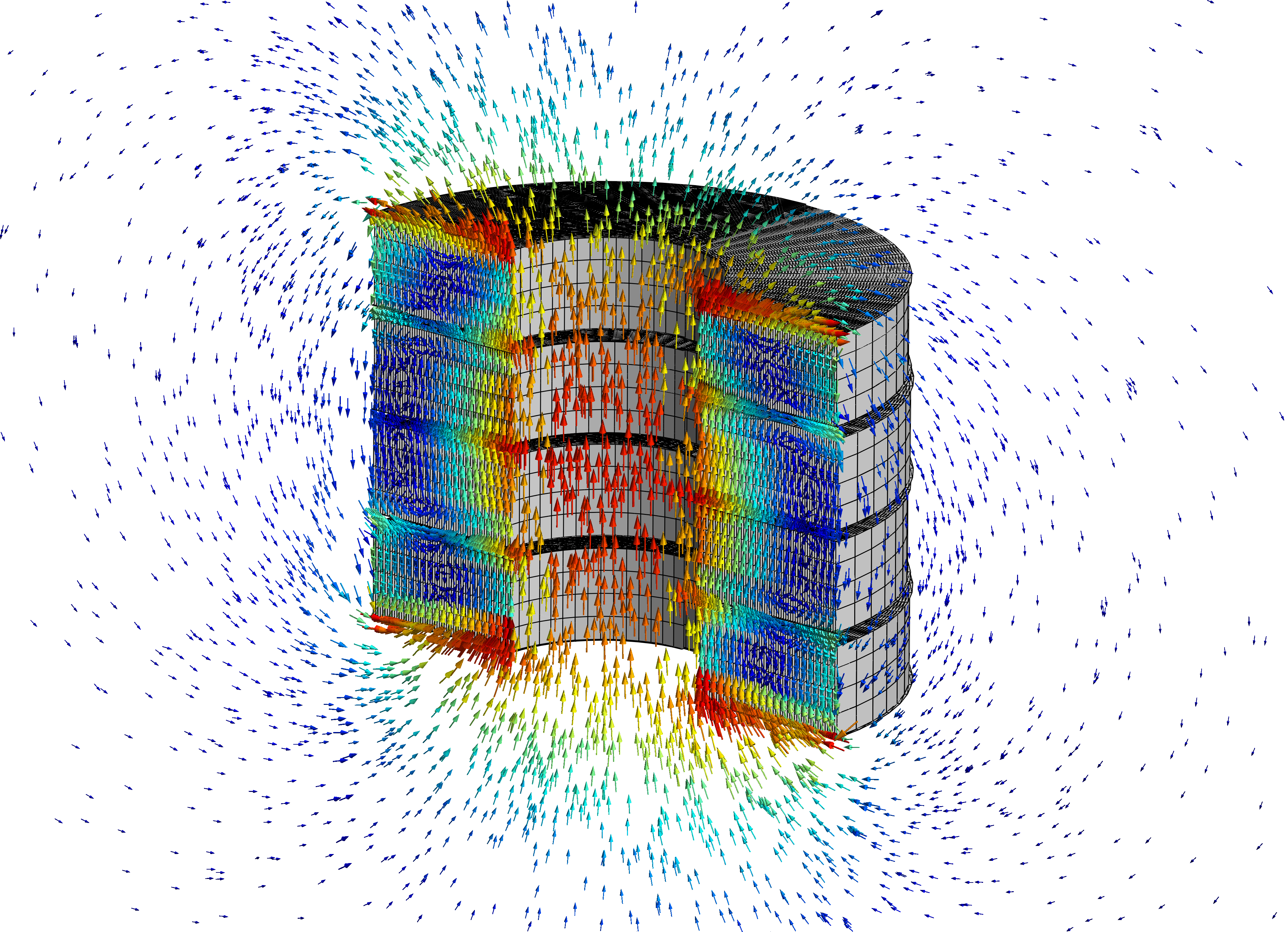}
    \caption{A clipped view of the magnetic field generated by a quadruple pancake coil with 60 turns per pancake, visualized in Gmsh's graphical user interface. Only the mesh of the windings is shown for clarity.}
    \label{fig:gmsh_gui_showcase}
\end{figure}

Finally, because GetDP can save the physical quantities in Gmsh format, Gmsh's graphical user interface and its postprocessing capabilities can be used to visualize the results. A clipped view of the magnetic field of a quadruple pancake coil is shown in \fref{fig:gmsh_gui_showcase}, which is visualized by Gmsh.

\tikzset{ 
    trim axis left,
    trim axis right
}
\section{Results}\label{sec:results}
This section presents a coupled magneto-thermal analysis of a double NI pancake coil with local degradation of the critical current density (local defect). Also, a parameter study is performed in \sref{sec:parameter_sweeps} where the simulation is repeated with ten different local defect lengths along the HTS CC of the spiral winding to understand the magneto-thermal characteristics of the model further. The results can be reproduced with \texttt{FiQuS} 2023.11.3, CERNGetDP 2023.11.0, and the provided input files \cite{referenceInputs}.

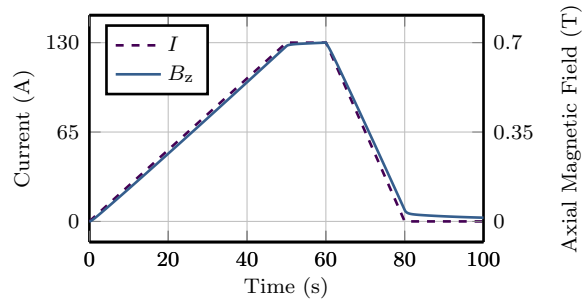
\begin{figure}[tbh]
    \centering
     \begin{tikzpicture}
	\begin{axis}[
        width=0.82\linewidth,
        height=\pgfPlotHeight,
    	xlabel=\pgfPlotXLabel,
        xtick=\pgfPlotXTicks,
    	ylabel={Current (A)}, 
        ytick={0, 65, 130}, 
    	axis y line*=left,
        ymax=150
     ]
        \pgfplotstableread[]{data/current_source.txt}{\dataCurrentSource}
        \addplot+[
            mark=none,
            index of colormap=0 of viridis,
            dashed
        ] table[x index={0}, y index={1}, col sep=comma, skip first n = 1]{\dataCurrentSource};
        \label{current}
	\end{axis}

	\begin{axis}[
        width=0.82\linewidth,
        height=\pgfPlotHeight,
        axis y line*=right,
    	ylabel={Axial Magnetic Field (T)}, 
        ytick={0, 0.35, 0.7}, 
        legend cell align={left},
        legend style={
            at={(0.04,0.96)},
            anchor=north west
        },
        grid=none,
        ymax=0.80769230769, 
    ]
        \addlegendimage{/pgfplots/refstyle=current}\addlegendentry{$I$}
        \pgfplotstableread[]{data/central_axial_magnetic_field.txt}\dataCentralAxialMagneticField
        \addplot+[
            mark=none,
            y filter/.code={\pgfmathparse{\pgfmathresult*(1)}\pgfmathresult},
            index of colormap=5 of viridis,
            solid
        ] table[x index = {1}, y index = {5}]{\dataCentralAxialMagneticField};
        \addlegendentry{$B_\text{z}$}
    \end{axis}
\end{tikzpicture}
    \caption{The current $I$ provided by the power supply, along with the axial magnetic field at the centre $B_z$ generated by the pancake coils.}
    \label{fig:current_and_central_axial_magnetic_field}
\end{figure}

\begin{figure}[tbh]
    \centering
    \includegraphics[width=0.7\linewidth]{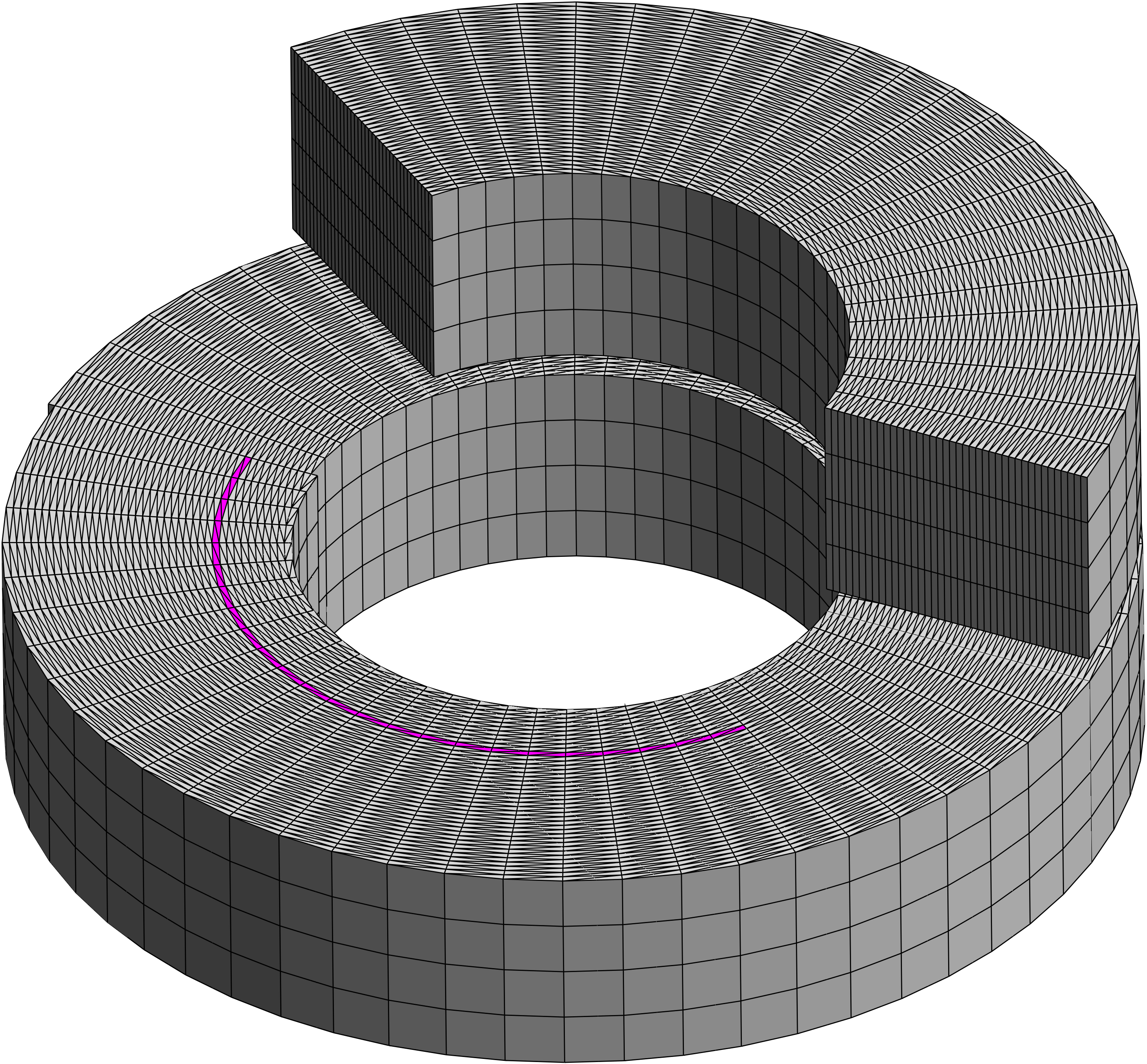}
    \caption{Clipped view of the double pancake coil windings' mesh with the local defect from turn 10.0 to 10.4 of the first pancake coil where the critical current density is set to \SI{0}{\ampere\per\metre\squared} at \SI{60}{\second}.}
    \label{fig:local_defect}
\end{figure}

All parameters regarding the geometry, material properties, and simulation setup are given in \tref{tab:results_summary}. All the material properties mentioned in the table and the critical current scaling function of ReBCO (\textit{HTS\_SuperPower}), $I_c(\vec{B},\,T)$, are taken from the STEAM material library \cite{steamMaterialLibrary}. The pancake coils are excited in current-driven mode. Firstly, the current is ramped from \SI{0}{\ampere} to \SI{130}{\ampere} in \SI{50}{\second}, followed by a plateau for \SI{10}{\second}, as shown in \fref{fig:current_and_central_axial_magnetic_field}. Then, a local defect, where the critical current density is set to \SI{0}{\ampere\per\metre\squared} from turn 10.0 to 10.4 of the first pancake coil, is introduced at \SI{60}{\second}. The local defect is depicted in \fref{fig:local_defect}. \SI{0.03}{\second} after the local defect, the current is ramped down from \SI{130}{\ampere} to \SI{0}{\ampere} in \SI{30}{\second}. The simulation is performed until \SI{100}{s}. The thin-shell approximation is used for the contact layer. For cooling conditions, the temperatures of the top and bottom surfaces of the terminals are fixed to \SI{4}{\kelvin}.

\begin{table}[tbh]
    \caption{Summary of the simulated model's parameters.}
    \label{tab:results_summary}
    \begin{indented}
        \centering
        \item[]\begin{tabular}{@{}ll}
            \br
            Description & Value\\
            \mr
            Number of coils & 2 \\
            Gap between the coils & \SI{0.5}{\milli\meter} \\
            Inner radius of windings & \SI{5}{\milli\meter} \\
            Number of turns for each coil & 40 \\
            HTS CC's thickness & \SI{120}{\micro\meter} \\
            HTS CC's width & \SI{4}{\milli\meter} \\
            Contact layer thickness & \SI{10}{\micro\meter} \\
            Inner terminal tube thickness & \SI{1}{\milli\meter} \\
            Outer terminal tube thickness & \SI{1}{\milli\meter} \\
            HTS CC: ReBCO thickness & \SI{1.5}{\micro\meter} \\
            HTS CC: Hastelloy\textsuperscript{\tiny\textregistered} thickness & \SI{75}{\micro\meter} \\
            HTS CC: Cu thickness & \SI{21}{\micro\meter} \\
            HTS CC: Cu RRR & 100 \\
            HTS CC: Ag thickness & \SI{1.5}{\micro\meter} \\
            HTS CC: Ag RRR & 100 \\
            HTS CC: Shunt layer (narrow side) & Cu, \SI{42}{\micro\meter} \\
            ReBCO Power law n-value & 30 \\
            $I_c(\vec{B}=\vec{0},\,T=\SI{4}{\kelvin})$ & \SI{770}{\ampere} \\
            ReBCO critical current scaling& $I_c(\vec{B},\,T)$ \\
            Contact layer's resistivity & \SI{1.12E-4}{\ohm\meter} \\
            Contact layer's material & Stainless steel \\
            Terminals' material & Copper \\
            Source current & $I(t)$ (\fref{fig:current_and_central_axial_magnetic_field}) \\
            Cooling conditions & $T=\SI{4}{\kelvin}$ (terminals)  \\
            Initial temperature & \SI{4}{\kelvin} \\
            Local defect ($I_c=0$) & \Fref{fig:local_defect}\\
            \br
        \end{tabular}
    \end{indented}
\end{table}

\begin{figure}
    \centering
    \begin{tikzpicture}
    \begin{axis}[
        width=1.14\pgfPlotWidth,
        axis equal image,
        hide axis,
        enlargelimits=false,
        point meta min = 0,
        point meta max = 230000000,
        colormap name=viridis,
        colorbar horizontal,
        colorbar style={
            title=Current Density $\left(\si{\ampere\per\meter\squared}\right)$,
            title style={
                at={(0.5,-1.35)},
                anchor=north,
                font=\footnotesize
            },
            scaled x ticks=false,
            xtick style={draw=none, font=\footnotesize},
            xtick={0, 77000000, 150000000, 230000000},
            at={(0,-0.1)},
            anchor=north west,
        },
    ] 
        \addplot graphics[xmin=0,xmax=1238,ymin=0,ymax=1049] {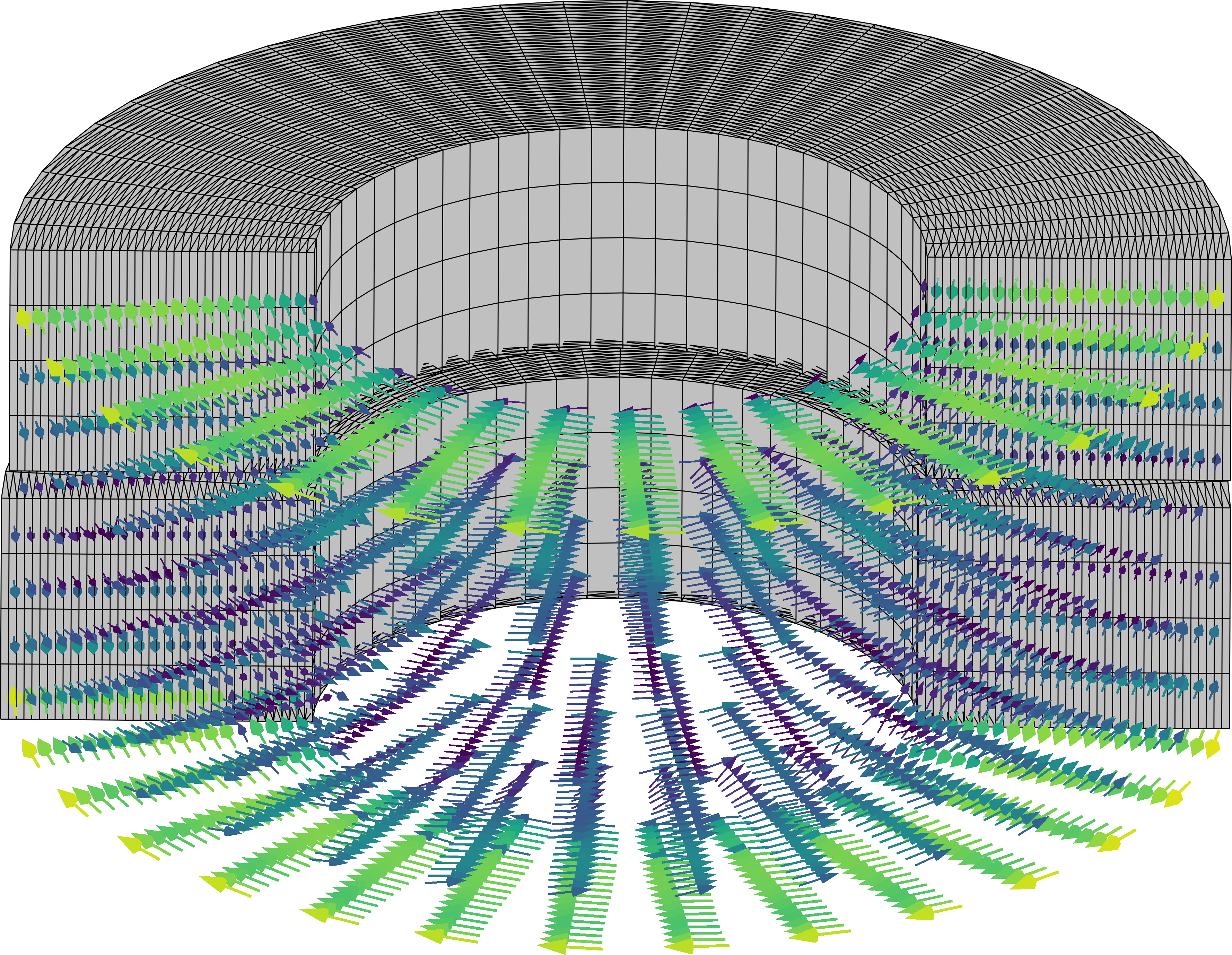};
    \end{axis}
\end{tikzpicture}%
    \caption{Clipped view of the current distribution in the coils at \SI{80}{\second}.}
    \label{fig:current_distribution}
\end{figure}

\Fref{fig:current_and_central_axial_magnetic_field} shows both the axial component of the central magnetic field $B_z$ and the source current $I$ versus time. The magnetic field is delayed compared to the source current due to the coils' inductance and radial currents. Moreover, although the source current decays to 0 at \SI{80}{\second}, $B_z$ still remains non-zero due to screening current-induced field \cite{Yang2013}. The current distribution at \SI{80}{\second} is shown in \fref{fig:current_distribution}, where screening currents are visible.

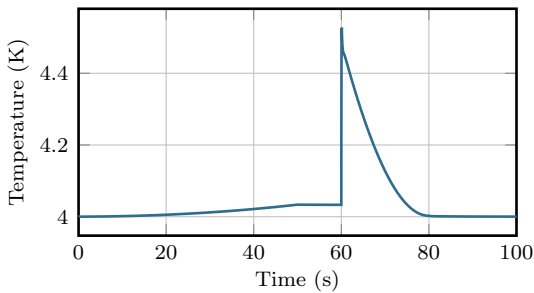
\begin{figure}[tbh]
    \centering
     \begin{tikzpicture}
	\begin{axis}[
        width=\pgfPlotWidth,
        height=\pgfPlotHeight,
    	xlabel=\pgfPlotXLabel,
        xtick=\pgfPlotXTicks,
    	ylabel={Temperature (K)}
     ]
        \pgfplotstableread[]{data/temperature.txt}\dataTemperature
        \addplot+[
            mark=none,
            index of colormap=6 of viridis,
            solid
        ] table[x index={1}, y index={5}, col sep=comma, skip first n = 1]{\dataTemperature};
    \end{axis}
\end{tikzpicture}
    \caption{Temperature at first pancake coil's turn 10.2 (the centre of the local defect).}
    \label{fig:temperature}
\end{figure}

\begin{figure}
    \centering
    \begin{tikzpicture}
    \centering
    \begin{axis}[
        width=1.14\pgfPlotWidth,
        axis equal image,
        hide axis,
        enlargelimits=false,
        point meta min = 4,
        point meta max = 4.4,
        colormap name=viridis,
        colorbar horizontal,
        colorbar style={
            title=Temperature (K),
            title style={
                at={(0.5,-1.35)},
                anchor=north,
                font=\footnotesize
            },
            xtick style={draw=none, font=\footnotesize},
            xtick={4, 4.1, 4.2, 4.3, 4.4},
            at={(0,-0.1)},
            anchor=north west
        },
    ] 
        \addplot graphics[xmin=0,xmax=1155,ymin=0,ymax=1082] {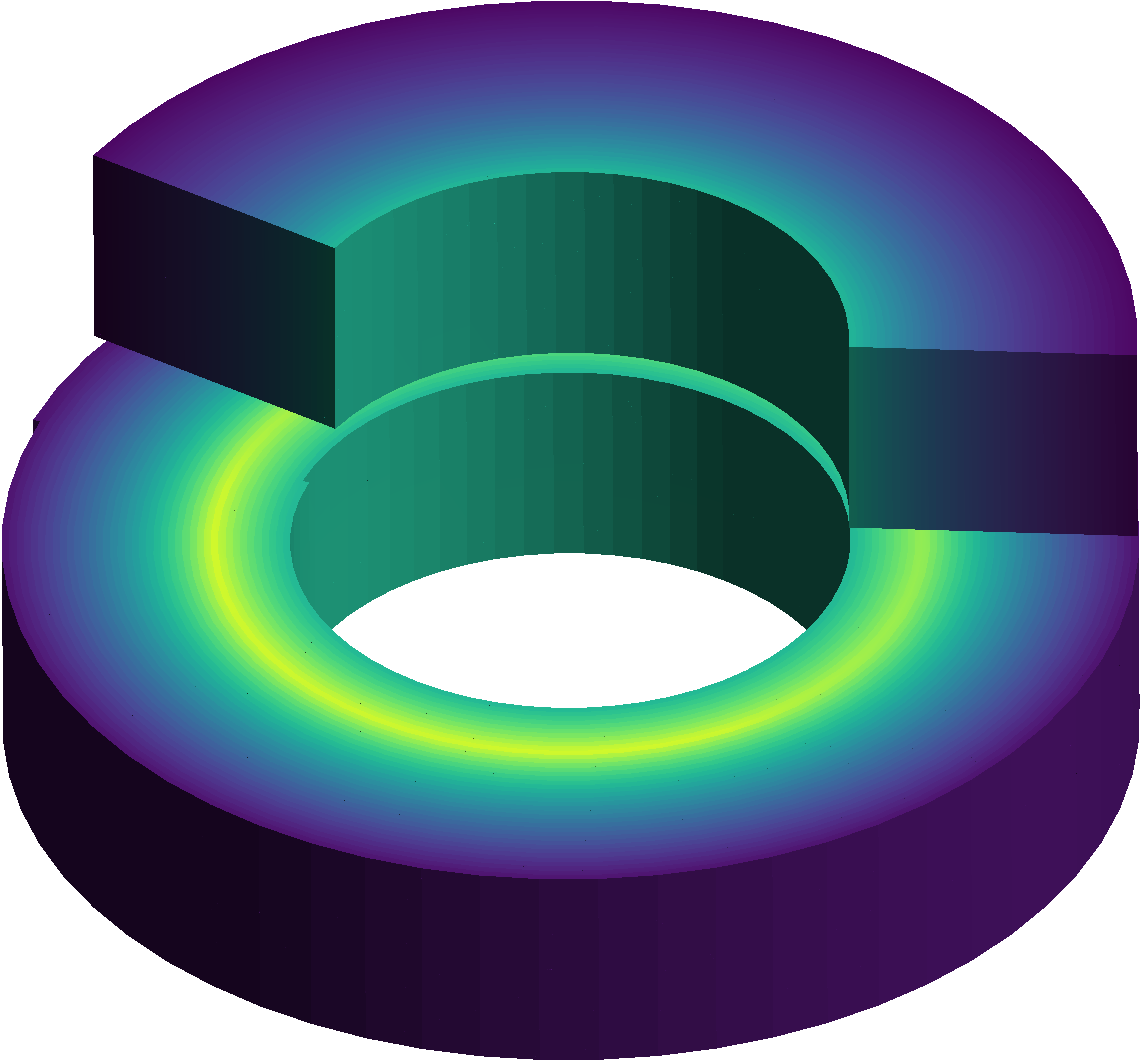};
    \end{axis}
\end{tikzpicture}%
    \caption{Clipped view of the temperature distribution in the coils at \SI{60.03}{\second}.}
    \label{fig:temperature_distribution}
\end{figure}

\Fref{fig:temperature} shows the temperature over time at the first pancake coil's turn 10.2, the centre of the local defect. During the coil ramp-up, temperature increases slightly due to resistive heating generated by radial currents \cite{Wang2017}. As the top and bottom surfaces of the terminals are kept at a fixed temperature with unlimited cooling power, this temperature increase is relatively small. Then, the temperature increases sharply at \SI{60}{\second} due to the sudden transition of turn 10.0 to 10.4 of the first pancake coil into a normal state at \SI{60}{\second}. The transition results in an increase in azimuthal resistivity locally, which causes a current redistribution and resistive heat generation. The generated heat is caused by high local azimuthal resistivity and contact layer resistivity since radial currents emerged. The temperature distribution at \SI{60.03}{\second} is shown in \fref{fig:temperature_distribution}, where the local hot spot around the local defect can be seen. Finally, the coil cools down slowly during ramping-down because of the cooling condition, the decreasing source current, and the stabilization of the current distribution.

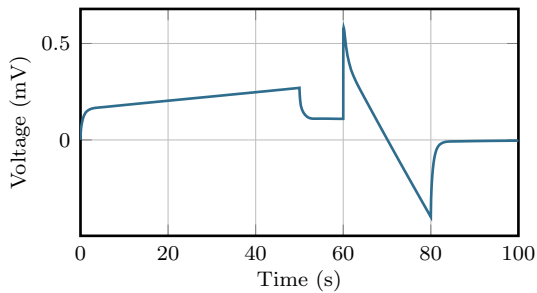
\begin{figure}[tbh]
    \centering
     \begin{tikzpicture}
	\begin{axis}[
        width=\pgfPlotWidth,
        height=\pgfPlotHeight,
    	xlabel=\pgfPlotXLabel,
        xtick=\pgfPlotXTicks,
    	ylabel={Voltage (mV)}, 
     ]
        \pgfplotstableread[]{data/voltage.txt}\dataVoltage
        \addplot+[
            y filter/.code={\pgfmathparse{\pgfmathresult*(1000)}\pgfmathresult},
            mark=none,
            index of colormap=6 of viridis,
            solid
        ] table[x index={0}, y index={1}, col sep=comma, skip first n = 1]{\dataVoltage};
    \end{axis}
\end{tikzpicture}
    \caption{Voltage difference between the terminals.}
    \label{fig:voltage}
\end{figure}

\Fref{fig:voltage} shows the voltage difference between the terminals induced by the ramp-up, local defect, and ramp-down. First, the voltage increases sharply during ramp-up due to the coil's characteristic response. After the coil stabilizes its response to ramping up, the voltage continues to increase linearly due to the terminal resistance. When ramping-up is complete, the voltage becomes constant after the coil's transient response ends. The non-zero voltage difference comes from the copper terminals, as the coil is still fully superconducting. Then, at \SI{60}{\second}, the voltage increases suddenly due to the increased azimuthal resistivity in the local defect and the radial currents caused by it. Finally, a negative voltage is seen while ramping down. A higher voltage difference occurs during ramping down than ramping up because the ramp-down is faster, as shown in \fref{fig:current_and_central_axial_magnetic_field}.

\subsection{Parametric Study of Local Defect Length}\label{sec:parameter_sweeps}
\begin{figure}[tbh]
    \centering
     \begin{tikzpicture}
	\begin{axis}[
        width=\pgfPlotWidth,
        height=1.1\pgfPlotSweepHeight,
    	xlabel=\pgfPlotXLabel,
        xtick={60, 70, 80},
    	ylabel={Temperature (K)}, 
        legend columns=1,
     ]
        \pgfplotstableread[]{data/temperature_sweep1.0.txt}\dataTemperatureSweepA
        \pgfplotstableread[]{data/temperature_sweep0.9.txt}\dataTemperatureSweepB
        \pgfplotstableread[]{data/temperature_sweep0.8.txt}\dataTemperatureSweepC
        \pgfplotstableread[]{data/temperature_sweep0.7.txt}\dataTemperatureSweepD
        \pgfplotstableread[]{data/temperature_sweep0.6.txt}\dataTemperatureSweepE
        \pgfplotstableread[]{data/temperature_sweep0.5.txt}\dataTemperatureSweepF
        \pgfplotstableread[]{data/temperature_sweep0.4.txt}\dataTemperatureSweepG
        \pgfplotstableread[]{data/temperature_sweep0.3.txt}\dataTemperatureSweepH
        \pgfplotstableread[]{data/temperature_sweep0.2.txt}\dataTemperatureSweepJ
        \pgfplotstableread[]{data/temperature_sweep0.1.txt}\dataTemperatureSweepK
        \addplot+[
            mark=none,
        ] table[x index={1}, y index={5}, col sep=comma, skip first n = 1]{\dataTemperatureSweepA};
        \addplot+[
            mark=none,
        ] table[x index={1}, y index={5}, col sep=comma, skip first n = 1]{\dataTemperatureSweepB};
        \addplot+[
            mark=none,
        ] table[x index={1}, y index={5}, col sep=comma, skip first n = 1]{\dataTemperatureSweepC};
        \addplot+[
            mark=none,
        ] table[x index={1}, y index={5}, col sep=comma, skip first n = 1]{\dataTemperatureSweepD};
        \addplot+[
            mark=none,
        ] table[x index={1}, y index={5}, col sep=comma, skip first n = 1]{\dataTemperatureSweepE};
        \addplot+[
            mark=none,
        ] table[x index={1}, y index={5}, col sep=comma, skip first n = 1]{\dataTemperatureSweepF};
        \addplot+[
            mark=none,
        ] table[x index={1}, y index={5}, col sep=comma, skip first n = 1]{\dataTemperatureSweepG};
        \addplot+[
            mark=none,
        ] table[x index={1}, y index={5}, col sep=comma, skip first n = 1]{\dataTemperatureSweepH};
        \addplot+[
            mark=none,
        ] table[x index={1}, y index={5}, col sep=comma, skip first n = 1]{\dataTemperatureSweepJ};
        \addplot+[
            mark=none,
        ] table[x index={1}, y index={5}, col sep=comma, skip first n = 1]{\dataTemperatureSweepK};

        \legend{
            $l$=1.0,
            $l$=0.9,
            $l$=0.8,
            $l$=0.7,
            $l$=0.6,
            $l$=0.5,
            $l$=0.4,
            $l$=0.3,
            $l$=0.2,
            $l$=0.1
        }
	\end{axis}
\end{tikzpicture}
    \caption{Temperature at the middle of the local defect for different local defect lengths $l$ in turn numbers.}
    \label{fig:sweep_temperature}
\end{figure}
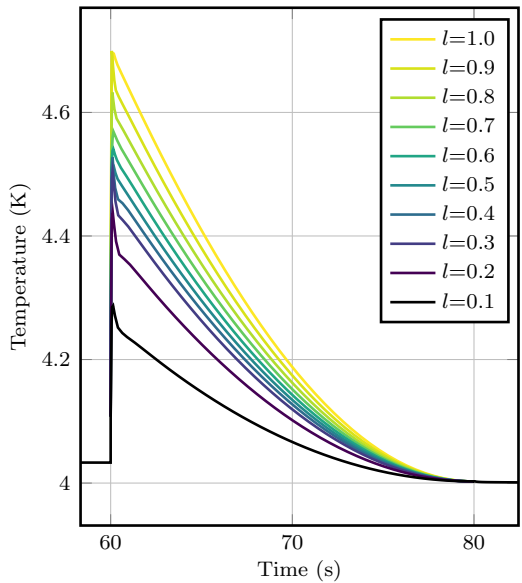

Because \texttt{FiQuS} is a part of the STEAM framework, parameter sweeps can be performed with the AnalysisSTEAM module of STEAM-SDK \cite{steamSDK}. Also, users can write their own Python script to sweep the parameters since \texttt{FiQuS} is distributed as a Python package \cite{fiqusPYPI}. With this capability, a parameter sweep study is conducted to further understand the effect of the local defect length along HTS CC on the results. Ten different lengths are considered. All these local defects start at turn 10 of the first pancake coil and end at 10.1, 10.2, ..., and 11.0. The voltage difference between the terminals and the temperature at the middle of the local defects are plotted over time in \fref{fig:sweep_voltage} and \fref{fig:sweep_temperature}, respectively. 

As expected, higher temperature and voltage increases are observed as the local defect length increases. At \SI{80}{\second}, the voltage differences between the terminals converge to the same value as they are caused by the coil's characteristic response to the ramping-down, and the local defect has a minimal effect on it.

\begin{figure}[tbh]
    \centering
     \begin{tikzpicture}
	\begin{axis}[
        width=\pgfPlotWidth,
        height=1.1\pgfPlotSweepHeight,
    	xlabel=\pgfPlotXLabel,
        xtick={60, 70, 80},
    	ylabel={Voltage (mV)}, 
        legend columns=1,
     ]
        \pgfplotstableread[]{data/voltage_sweep1.0.txt}\dataVoltageSweepA
        \pgfplotstableread[]{data/voltage_sweep0.9.txt}\dataVoltageSweepB
        \pgfplotstableread[]{data/voltage_sweep0.8.txt}\dataVoltageSweepC
        \pgfplotstableread[]{data/voltage_sweep0.7.txt}\dataVoltageSweepD
        \pgfplotstableread[]{data/voltage_sweep0.6.txt}\dataVoltageSweepE
        \pgfplotstableread[]{data/voltage_sweep0.5.txt}\dataVoltageSweepF
        \pgfplotstableread[]{data/voltage_sweep0.4.txt}\dataVoltageSweepG
        \pgfplotstableread[]{data/voltage_sweep0.3.txt}\dataVoltageSweepH
        \pgfplotstableread[]{data/voltage_sweep0.2.txt}\dataVoltageSweepJ
        \pgfplotstableread[]{data/voltage_sweep0.1.txt}\dataVoltageSweepK
        
        \addplot+[
            y filter/.code={\pgfmathparse{\pgfmathresult*(1000)}\pgfmathresult},
            mark=none
        ] table[x index={0}, y index={1}, col sep=comma, skip first n = 1]{\dataVoltageSweepA};
        \addplot+[
            y filter/.code={\pgfmathparse{\pgfmathresult*(1000)}\pgfmathresult},
            mark=none
        ] table[x index={0}, y index={1}, col sep=comma, skip first n = 1]{\dataVoltageSweepB};
        \addplot+[
            y filter/.code={\pgfmathparse{\pgfmathresult*(1000)}\pgfmathresult},
            mark=none
        ] table[x index={0}, y index={1}, col sep=comma, skip first n = 1]{\dataVoltageSweepC};
        \addplot+[
            y filter/.code={\pgfmathparse{\pgfmathresult*(1000)}\pgfmathresult},
            mark=none
        ] table[x index={0}, y index={1}, col sep=comma, skip first n = 1]{\dataVoltageSweepD};
        \addplot+[
            y filter/.code={\pgfmathparse{\pgfmathresult*(1000)}\pgfmathresult},
            mark=none
        ] table[x index={0}, y index={1}, col sep=comma, skip first n = 1]{\dataVoltageSweepE};
        \addplot+[
            y filter/.code={\pgfmathparse{\pgfmathresult*(1000)}\pgfmathresult},
            mark=none
        ] table[x index={0}, y index={1}, col sep=comma, skip first n = 1]{\dataVoltageSweepF};
        \addplot+[
            y filter/.code={\pgfmathparse{\pgfmathresult*(1000)}\pgfmathresult},
            mark=none
        ] table[x index={0}, y index={1}, col sep=comma, skip first n = 1]{\dataVoltageSweepG};
        \addplot+[
            y filter/.code={\pgfmathparse{\pgfmathresult*(1000)}\pgfmathresult},
            mark=none
        ] table[x index={0}, y index={1}, col sep=comma, skip first n = 1]{\dataVoltageSweepH};
        \addplot+[
            y filter/.code={\pgfmathparse{\pgfmathresult*(1000)}\pgfmathresult},
            mark=none
        ] table[x index={0}, y index={1}, col sep=comma, skip first n = 1]{\dataVoltageSweepJ};
        \addplot+[
            y filter/.code={\pgfmathparse{\pgfmathresult*(1000)}\pgfmathresult},
            mark=none
        ] table[x index={0}, y index={1}, col sep=comma, skip first n = 1]{\dataVoltageSweepK};

        \legend{
            $l$=1.0,
            $l$=0.9,
            $l$=0.8,
            $l$=0.7,
            $l$=0.6,
            $l$=0.5,
            $l$=0.4,
            $l$=0.3,
            $l$=0.2,
            $l$=0.1
        }
	\end{axis}
\end{tikzpicture}
    \caption{Voltage difference between the terminals obtained for different local defect lengths $l$ in turn numbers.}
    \label{fig:sweep_voltage}
\end{figure}
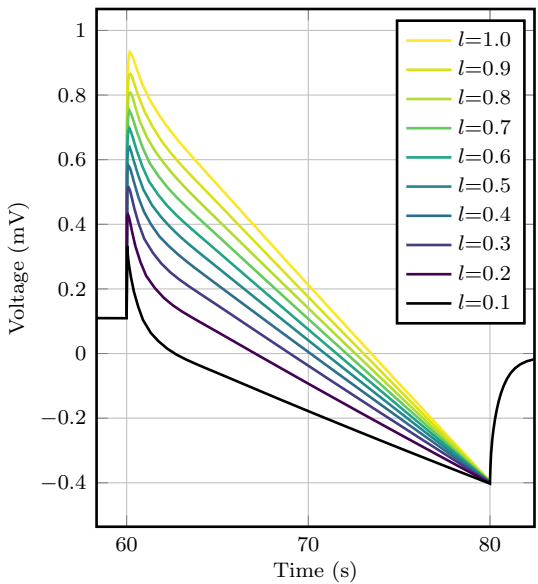

\section{Conclusions}\label{sec:conclusions}
In this paper, the new \texttt{Pancake3D} module of the open-source and parametric FE quench simulation tool \texttt{FiQuS} has been presented. \texttt{FiQuS}/\texttt{Pancake3D} is a Python tool that conducts 3D FE quench simulations of NI pancake coils. \texttt{FiQuS} uses a text-based YAML input file that fully specifies the model. The structure of the input file is designed to make it easy and intuitive to use, regardless of the users' level of expertise in modelling and simulation. \texttt{FiQuS}/\texttt{Pancake3D} is capable of generating geometry, generating mesh, solving, and postprocessing an FE-based coupled magneto-thermal or magnetodynamic simulation of the transient behaviour of NI pancake coils.

\texttt{FiQuS}/\texttt{Pancake3D} addresses common challenges of FE simulation of NI coils. The functionalities and capabilities of \texttt{FiQuS}/\texttt{Pancake3D}'s geometry generator, mesh generator, solver, and postprocessor have been explained, along with examples and technical strategies to develop these modules. As a showcase of these capabilities, the results of a transient magneto-thermal simulation of a double pancake coil with 40 turns are presented. Finally, with \texttt{FiQuS}/\texttt{Pancake3D}'s parametric capabilities, a parameter sweep study is conducted where the same simulation is repeated with different local defect lengths.

\texttt{FiQuS} is an ongoing collaborative project that is actively developed at CERN. Ongoing work focuses on incorporating enhancements and supporting new types of magnets and conductors.

\section*{Acknowledgment}
The work of E. Schnaubelt is supported by the Graduate School CE within the Centre for Computational Engineering at TU Darmstadt and by the Wolfgang Gentner Programme of the German Federal Ministry of Education and Research (grant no. 13E18CHA).

\printbibliography

\end{document}